# Deep Learning-based Unsupervised Domain Adaptation via a Unified Model for Prostate Lesion Detection Using Multisite Bi-parametric MRI Datasets

Hao Li, PhD[1,3] • Han Liu, PhD[1,3] • Heinrich von Busch, PhD[2] • Robert Grimm, PhD[2] • Henkjan Huisman, PhD[4] • Angela Tong, MD[5] • David Winkel, MD[6] • Tobias Penzkofer, MD[7] • Ivan Shabunin, MD[8] • Moon Hyung Choi, MD[9] • Qingsong Yang, MD[10] • Dieter Szolar, MD[11] • Steven Shea, MD[12] • Fergus Coakley, MD[13] • Mukesh Harisinghani, MD[14] • Ipek Oguz, PhD[3] • Dorin Comaniciu, PhD[1] • Ali Kamen, PhD[1] • Bin Lou, PhD[1]

[1] Digital Technology and Innovation, Siemens Healthineers, Princeton, NJ, USA

[2] Diagnostic Imaging, Siemens Healthineers, Erlangen, Bavaria, Germany

[3] Vanderbilt University, Nashville, TN, USA

[4] Radboud University Medical Center, Nijmegen, NL

[5] New York University, New York City, NY, USA

[6] Universitätsspital Basel, Basel, Switzerland

[7] Charité, Universitätsmedizin Berlin, Berlin, Germany

[8] Patero Clinic, Moscow, Russia

[9] Eunpyeong St. Mary's Hospital, Catholic University of Korea, Seoul, Republic of Korea

[10] Radiology Department, Changhai Hospital of Shanghai, China

[11] Diagnostikum Graz Süd-West, Graz, Austria

[12] Department of Radiology, Loyola University Medical Center, Maywood, IL, USA

[13] Diagnostic Radiology, School of Medicine, Oregon Health and Science University, Portland, OR, USA

[14] Massachusetts General Hospital, Boston, MA, USA

# Abstract

**Purpose:**

To determine whether the unsupervised domain adaptation (UDA) method with generated images improves the performance of a supervised learning (SL) model for multi-site prostate cancer (PCa) detection.

**Materials and Methods:**

This retrospective study included data from 5,150 patients (14,191 samples) collected across nine different imaging centers. A novel UDA method using a unified generative model was developed for multi-site PCa detection. This method translates diffusion-weighted imaging (DWI) acquisitions, including apparent diffusion coefficient (ADC) and individual DW images acquired using various b-values, to align with the style of images acquired using b-values recommended by Prostate Imaging Reporting and Data System (PI-RADS) guidelines. The generated ADC and DW images replace the original images for PCa detection. An independent set of 1,692 test cases (2,393 samples) was used for evaluation. The area under the receiver operating characteristic curve (AUC) was used as the primary metric, and statistical analysis was performed via bootstrapping.

**Results:**

For all test cases, the AUC values for baseline SL and UDA methods were 0.73 and 0.79 (p<.001), respectively, for PI-RADS≥3, and 0.77 and 0.80 (p<.001) for PI-RADS≥4 PCa lesions. In the 361 test cases under the most unfavorable image acquisition setting, the AUC values for baseline SL and UDA were 0.49 and 0.76 (p<.001) for PI-RADS≥3, and 0.50 and 0.77 (p<.001) for PI-RADS≥4 PCa lesions.

## Conclusion:

UDA with generated images improved the performance of SL methods in multi-site PCa lesion detection across datasets with various b values, especially for images acquired with significant deviations from the PI-RADS recommended DWI protocol (e.g. with an extremely high b-value).

**Summary:**

Unsupervised domain adaptation with diffusion-weighted images generated using a unified model improved the performance of supervised learning models in PCa lesion detection across large, multisite datasets with various b-values.

**Key Points:**

(1) A novel unsupervised domain adaptation (UDA) method used a unified generator to translate the diffusion-weighted image style from multiple target (test) domains into the reference (training) domain used for training the supervised learning prostate cancer detection model.

(2) The dynamic filter was used to enable multi-domain mapping within a unified generator that accepts target images with arbitrary b-values as distinct domains, leveraging meta-information to differentiate between domains.

(3) When evaluated on 1,692 (2,393 samples) unseen test cases, the proposed method was shown to improve performance of the baseline SL method in detecting PI-RADS $\geq 3$ lesions (area under the receiver operating characteristic curve (AUC), 0.73 to 0.79 p<.001), including when the b-values of the test image represented out-of-distribution samples from the training set (particularly for low-b= {150, 200} and high-b=2000, with the AUC increasing from 0.49 to 0.76 (p<.001)).

**Abbreviations**

PCa = prostate cancer, bp-MRI = bi-parametric MRI, DWI = diffusion-weighted imaging, ADC = apparent diffusion coefficient, UDA = unsupervised domain adaptation, PI-RADS = Prostate Imaging Reporting and Data System, AUC = area under the receiver operating characteristic curve, FROC = free-response receiver operating characteristic, SL = supervised learning.

# Main Body

## 1.	Introduction

Prostate cancer (PCa) is one of the most common cancers in men. If detected early the patient can have an improved prognosis, including better treatment outcomes and lower mortality rates (1, 2). Earlier studies have shown promising results for early PCa diagnosis using multiparametric MRI (3, 4) or biparametric MRI (bpMRI) (5, 6). Recently, deep learning-based methods have achieved high performance for PCa detection by leveraging information from bpMRI images (7, 8, 9, 10, 11). These methods could boost productivity of radiologists by shortening the time needed for interpreting imaging through automated lesion detection. Additionally, they have the potential to heighten diagnostic accuracy, notably for less experienced radiologists, and to enhance consistency among different readers (9). In these techniques, diffusion-weighted imaging (DWI) stands out as a crucial element, offering a pronounced distinction in signal intensities between cancerous and healthy tissues. This distinction is especially noticeable in apparent diffusion coefficient (ADC) maps and high b-value images.

Many convolutional neural networks (CNNs) from prior studies were trained and tested using supervised learning (SL) methods on datasets either from a single institution or multiple sites that adhered to similar acquisition protocols, particularly with b-value settings as recommended by the Prostate Imaging Reporting and Data System (PI-RADS (12)) guidelines. Under these conditions, test samples are tightly matched to the training set, allowing CNNs to yield reliable results for such in-distribution data. However, in real-world scenarios, clinical sites may have their own preferences for b-value selections. While variances in ADC and high b-value images due to diverse b-value choices may seem negligible to human observers, they can substantially influence deep learning models. This is largely attributed to the domain shifts observed across images from different datasets (13). The performance of a CNN drops, producing inaccurate results, when encountering domain shifts or processing an out-of-distribution (OOD) test sample from a target domain, whose b-values are not included in the training set (14). Fig. 1(a) provides an illustrative example.

One straightforward way to overcome the domain shift problem is to retrain the generic model with a larger dataset to enlarge the b-values distribution. Yet, in practice, it is hard to transfer data between different imaging centers to obtain extra training data, and obtaining human labels for large datasets is time-consuming and expensive. In addition, there is no assurance that all b-values will be included.

Domain adaptation is a potential solution to address the domain shift issue, which attempts to alleviate the decrease of generalization ability caused by the distribution shift between source domain training data and target domain test data (15, 16, 17, 18). However, such methods require human delineations from the target domain during the training process. Addressing the label availability challenge, unsupervised domain adaptation (UDA) has been introduced. While UDA methods are extensively employed across various medical image analysis tasks (19, 20, 21, 22), only a few studies have been conducted in the field of PCa detection (23). Furthermore, most existing UDA studies are focused on single-domain mapping. When applied to multi-domain settings, these methods are required to train multiple generators for every domain pair, as well as their downstream task networks, after obtaining the generated labeled data. This approach is time-consuming and not feasible in practice (Fig. 1(b)).

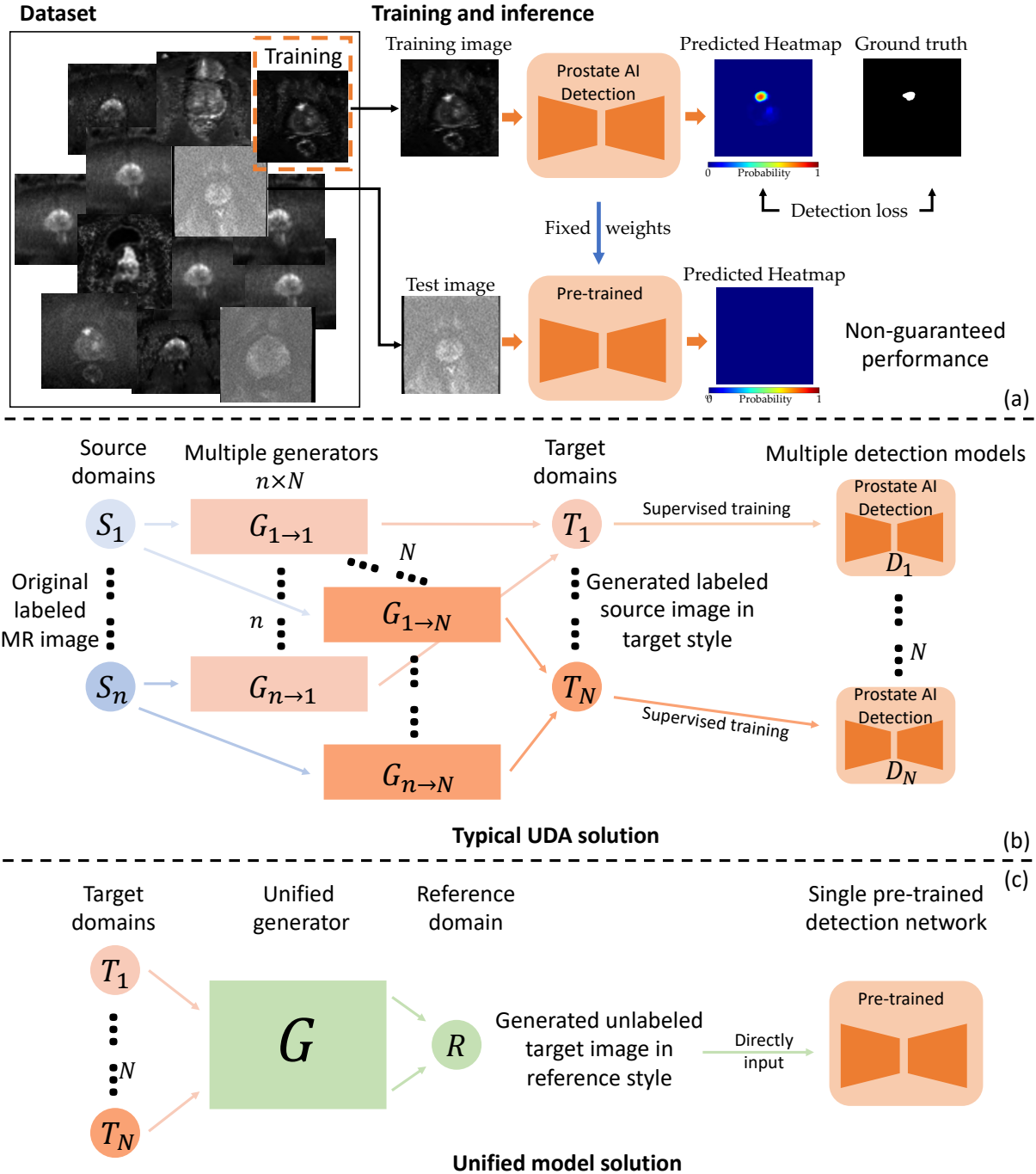

Figure 1: (a) Diagram shows the common domain shift problem for prostate cancer detection with a supervised learning model. When MR images are collected using very different protocols, the performance of the pre-trained model is not guaranteed. (b) Diagram shows the typical unsupervised domain adaptation (UDA) solution on domain shift, in which multiple generators and detection models are trained. (c) Our unified model solution, which needs only a single generator and pre-trained detection network. Specifically, the proposed method aims to translate image style from the target domain (unlabeled test data) to the reference domain (labeled data used to train the detection model) using a unified generator. S and T indicate source and target domains, respectively. R represents the reference domain. The training set is regarded as the source domain, with the reference domain being its subset. Best viewed in color. AI = artificial intelligence.

The aim of this study was to assess whether harmonization of ADC map and high b-value image data improve the accuracy of predicting prostate lesion from multi-site data with various b-values. Designed for practical applications, we propose a UDA framework with a unified model for multi-domain mapping, as depicted in Fig. 1(c), which is computationally efficient, especially when encountering a large number of domains. Moreover, our method does not require annotations from the test data, and it can be applied to any pre-trained network for practical use without retraining. Specifically, our approach uses the ADC map and high b-value images derived from various b-value settings as inputs. It leverages meta-information to more effectively synthesize the ADC map and high b-value image at a consistent (and standard) b-value setting. Our method was evaluated using a large-scale dataset with various b-values.

## 2. Materials and Methods

### 2.1 Dataset

This retrospective analysis uses a multi-cohort dataset of 5,150 patients from nine different clinical sites, all of which have bpMRI prostate examinations consisting of T2-weighted (T2w) acquisition and DWI of at least two different b-values. Of the 5,150 patients, 2,170 have been previously reported in (9). The inclusion criteria for the study were as follows: (1) patients who were treatment-naïve; (2) clear visibility of the prostate gland in the field of view of bi-parametric MRI images; (3) acquisition of images using 1.5 T or 3 T axial MRI scans with either a body or endorectal coil. Conversely, the exclusion criteria were: (1) cases involving prostatectomy or any scenario where the prostate was partially resected; and (2) cases with severe artifacts resulting from implants or motion. The prior article used only one pair of b-values for each case; whereas in this study, all qualified b-values were included to investigate their impact on the detection performance. The b-values ranging from 0 to 200 were considered as low b-values, while those between 600 and 2000 were considered as high b-values. The ADC images and DWI b=2000 images were computed based on each pair of low b-value and high b-value DWI images by performing a nonlinear least-squares fit to the equation $S(b) = S_0 \cdot e^{-b \cdot ADC}$. For each voxel, the coefficient of b was employed as its ADC value (with a scaling factor of $10^6$), and the intensity of B-2000 image was calculated through extrapolation at b=2000. To maintain consistency and reduce variation in ADC computation, vendor-provided ADC maps were excluded from the study. In this way, each pair of b values from the same case can be considered as a unique sample from a different domain. This yielded a total of 14,191 samples of 34 different combinations of b-values from all cases. We categorized all samples into a few subgroups based on the range of b-values. The details of each subgroup can be viewed in

Fig. 2(a).

A total of 3,458 cases were used for training, as shown in Fig. 2(b). For training the baseline method (i.e. the SL model from (24)), the best pair of b-values (optimal) was selected and only one single sample from each case was used. PI-RADS guidelines recommend using one low b-value set at 0-100 $sec/mm^2$ (preferably 50-100 $sec/mm^2$) and one intermediate b-value set at 800-1000 $sec/mm^2$ for ADC computations (25). We followed this suggestion and selected the b-values that are the closest to 50 and 1000 as low and high b-values, respectively. For other methods (generic model and proposed UDA methods), additional samples with all possible b-value pairs were used, which consisted of 11,763 samples from the same training cases. In the UDA training process, 882 samples whose b-values are from the standard domain (low b-value=50, high b-value=800) were selected as reference domain data to train the unified generator of UDA methods, and the rest of the data were considered target domain samples. The independent testing set contained 1,692 cases with 2,428 samples. The results of 2,393 samples are reported in this work due to very limited sample number for some b-value subgroups, such as groups 10 and 11 in Fig. 2(a). All the cases had lesion-based PI-RADS information and voxel-based annotations of the lesion boundaries. The PCa lesion annotations were obtained based on the clinical radiology reports and carefully reviewed by an expert radiologist (DW) with five years of experience in radiology, specializing in prostate MRI examinations; the details of annotation process can be viewed in S.1 (Supplementary Materials). A positive case was identified if it contained PI-RADS$\geq$ 3 lesions.

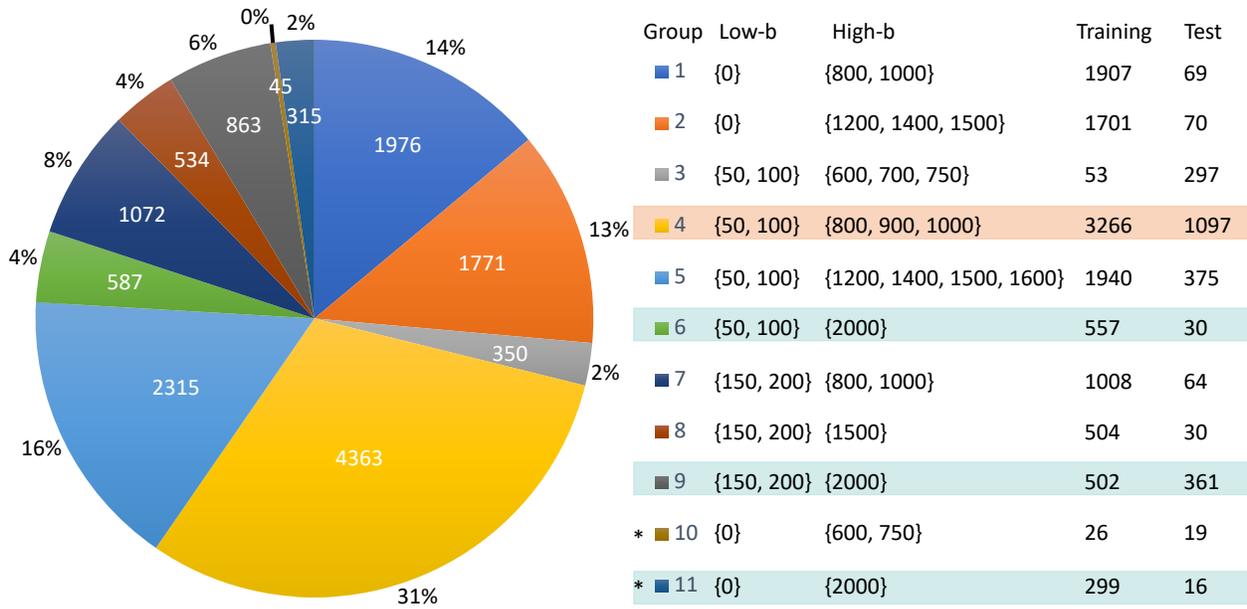
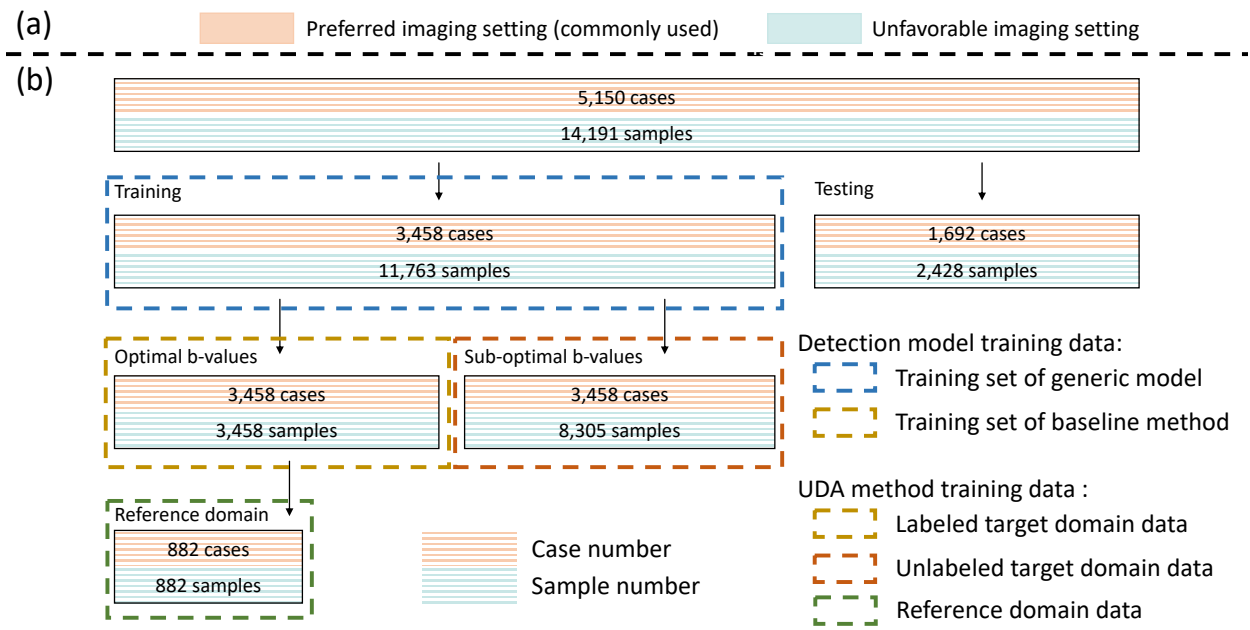

Figure 2: Overview of the dataset. The (a) b-value distribution and (b) data split are shown. The left side (pie chart) of (a) shows the sample number and percentage of each group, and the details are shown on the right side. * denotes the groups excluded from testing due to the limited sample number. In (b), the detailed training data for different methods are marked by different dashed rectangle boxes; best viewed in color.

## 2.2 Image Preprocessing

We adopted the same preprocessing procedures as described in (14, 24). PI-RADS guidelines recommends a high b-value set at $\geq$1,400 $sec/mm^2$; similarly, in this study, we re-computed a high b-value image at a fixed $b$ =2,000 $sec/mm^2$ to ensure a representation in which lesions stand out. The fixed b-value was selected to further eliminate the b-value variances among datasets (26). The preprocessing pipeline took raw bpMRI acquisitions and generated well-formatted data volumes for all subsequent synthesis and detection models (see S.2).

## 2.3 Algorithm Design

Fig. 3 shows the proposed framework which aims to solve two common practical issues in PCa detection, i.e. domain shift and label availability for test data. The proposed framework contains two parts: synthesis and detection. To increase the generalizability of the SL detection network for OOD test samples, generators align the style of DWI B-2000 and ADC test samples from the target domain to the reference domain at the image level. Next, the detection model predicts the PCa heatmap which uses the concatenation of T2w, generated ADC, generated DWI B-2000, and prostate mask as inputs. Notably, this entire process operates without the need for test data labels. To more accurately mimic real-world scenarios, we initially trained the detection model and then used the trained model to educate the generators.

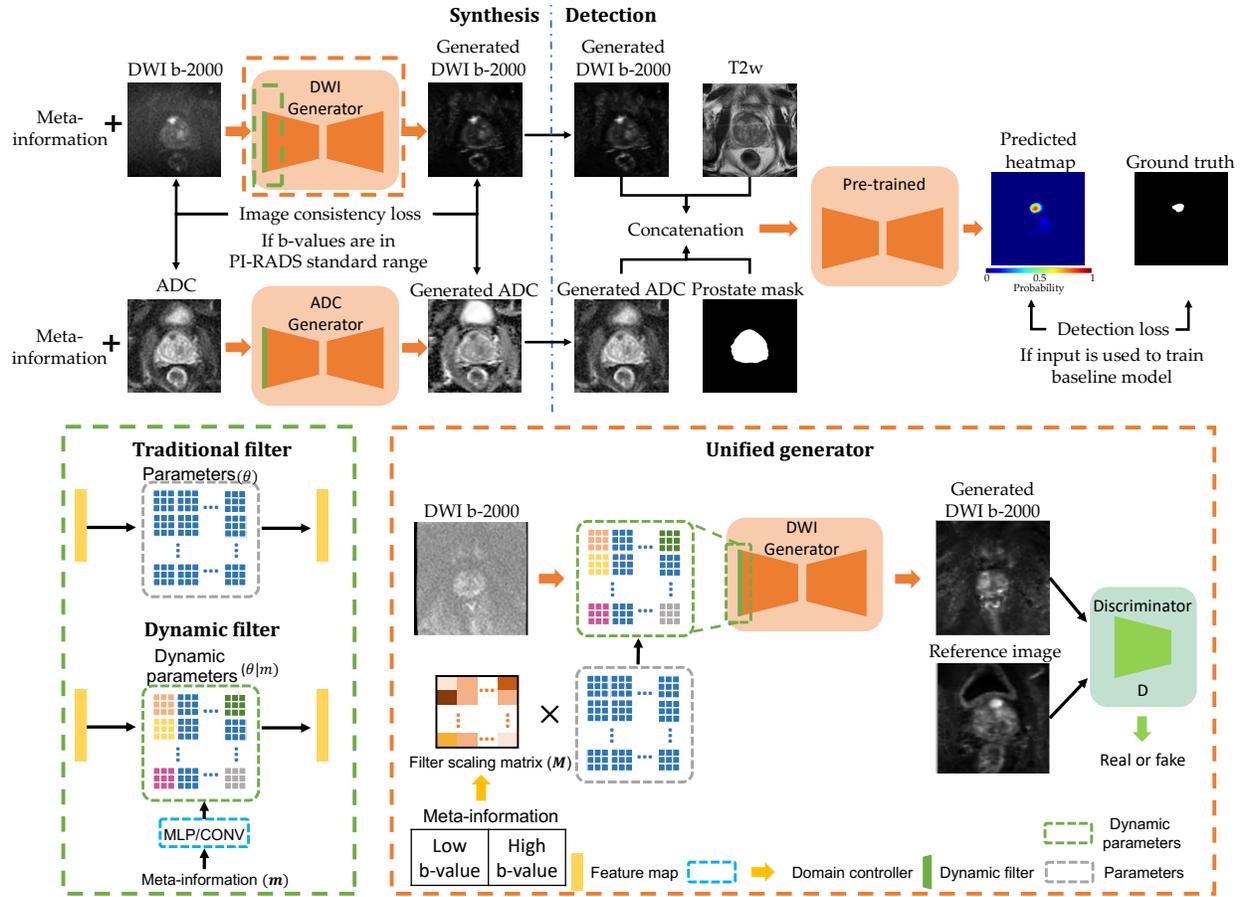

*Figure 3: Proposed unsupervised domain adaptation (UDA) framework for prostate cancer (PCa) detection, which contains two parts: synthesis (top left) and detection (top right). Specifically, the supervised learning (SL) model from baseline method (24) is first trained to provide supervision to the generators. In addition, both generators (orange) are identical but with different weights. The dynamic filter (green) is plugged into the generator for multi-domain mapping and takes meta-information as input to specify domain information. The bottom left shows the differences between traditional and dynamic filters, and the bottom right provides details of the proposed unified generator.*

## 2.4 Detection Network

Our 2D detection network is a U-Net embedded with residual blocks. This network uses information from the 2D slice of T2w, ADC, DWI B-2000, and prostate mask to generate the corresponding PCa lesion heatmap. Further details of the architecture can be referenced in (24). In line with the configurations of (24), a 3D heatmap is derived from all slices for each sample, with the non-zero regions considered as lesion candidates. To identify the true positives (TPs) and false positives (FPs), a threshold was used to obtain a set of connected components. The TPs were be identified if the connected components overlapped on annotations or were less than 5mm away from the lesion center. Otherwise, such connected components were classified as FPs. Any PCa lesions that lacked corresponding detections were termed false negatives (FNs).

**2.5 Synthesis Network and Dynamic Filter**

The synthesis models (generators) were adapted from CUT (Contrastive learning for Unpaired image-to-image Translation) (27), which is a U-shaped network. While the architecture for the DWI B-2000 and ADC generators is identical, they have different parameters (i.e., separate networks are trained for the DWI B-2000 and ADC images). Each generator takes a 2D image from the target domain as input and produces a 2D image styled in the reference domain, as illustrated in Fig. 3. The discriminator evaluates the performance of the generator during training by distinguishing between generated and real reference domain images. When both the input image and its label were used to train the baseline method, the detection loss was applied to offer supplementary guidance to the generator. To emphasize correct mapping to the reference domain, we employed an additional consistency loss ($L_{consistency}$) at the image level to preserve the original information from the input image, especially if its b-values aligned with the PI-RADS standard range. Specifically, the mean square error serves as the metric.

Unlike typical UDA methods, which require multiple networks for multi-domain mapping, we used only a unified model for each modality. However, the performance of the unified model may be limited in producing robust results between multiple domains due to a lack of domain information. To address this issue, we proposed a dynamic filter as a domain indicator, which aims to increase domain generalizability of the unified model by leveraging meta-information. As shown in Fig. 3, the traditional convolutional layer processes feature maps with parameters ($\theta$) that are learned during training. In contrast, the parameters of the dynamic filter are dynamically generated based on different conditions by a scaling layer. Thus, the unified generator could achieve robust multi-domain mapping with provided meta-information. Other works use one-hot encoding for the dynamic filter (28, 29, 30), which can be complex for large-scale studies with multiple domains. In contrast, we converted low and high b-values from meta-information into a 2D tensor, which simplifies the input for the domain controller and preserves original information. As suggested in (29, 31), a filter scaling strategy generates a kernel-wise scale factor, uniformly weighting all parameters instead of individually scaling them. Specifically, the domain controller learns to generate the corresponding filter scaling matrix ($M$) based on the provided meta-information. Each element in $M$ represents the scale of the corresponding kernel, and the parameters of the dynamic filter are dynamically adjusted through scalar multiplication.

## 2.6 Implementation Details

The training process of the detection model (baseline) is detailed in (24), and we used binary cross-entropy as detection loss ($L_{det}$). To train the generator, we set the batch size to 96 and used the same loss functions as (27), which are denoted as $L_{CUT}$. The total epoch was set to 100. The domain controller is a simple convolutional layer with 7 as kernel size, an input channel of 2, and an output channel of 128, where 64 scaling factors and bias weights are included. The hyperparameters of the generators are the same as (27). In addition, the loss selections of generators ($L_{syn}$) depend on three scenarios related to the input image: (1) $L_{syn} = L_{CUT}$ for unlabeled target domain data; (2) $L_{syn} = L_{det} + L_{CUT}$ for labeled target domain data; and (3) $L_{syn} = L_{det} + L_{CUT} + L_{consistency}$ for reference domain data. The detailed training process of generator can be viewed in S.3. The training was conducted on NVIDIA A100 GPUs and implemented using PyTorch.

## 2.7 Model Comparisons

We compared our proposed framework with two deep learning methods for multi-site PCa lesion detection, which are (1) baseline: a SL pre-trained detection model (24); and (2) generic model: retrain the baseline method using a larger dataset with various b-values. We also reported the results of the ablation study to show the effectiveness of the proposed method in S.6.

## 2.8 Statistical Analysis

The area under the receiver operating characteristic curve (AUC) score was computed as case-level performance, which is the primary endpoint of this work. The maximum value of the 3D heatmap was defined as the prediction score of the sample to calculate the AUC score. The confidence interval was computed based on a bootstrap approach with 2,000 resamples. The statistical results were conducted with Python with numpy, sklearn and scipy libraries We set a statistical significance threshold of 0.05. In addition, the free-response receiver operating characteristic (FROC) was used as a metric to evaluate the lesion-level performance as supplementary results. Moreover, peak signal-to-noise ratio (PSNR), mean square error (MSE), and structural similarity index measure (SSIM) were used as metrics to evaluate the image quality of generated images. This requires the same cases to have DWI images in both reference domain and other target domains. To achieve this, we selected all cases from the testing dataset that had six different b-values. The DWI B-2000 images were computed using naturally acquired DWI images of three different b-value pairs: (50, 800), (150, 1500) and (200, 2000). The proposed method was applied to the B-2000 images computed using b-values of (150, 1500) and (200, 2000) to generate new B-2000 images. The original and generated B-2000 images were compared with the corresponding one computed by using (50, 800) b-values. We used t-SNE visualization to assess the impact of the generated ADC and DWI B-2000 images on the detection network. Specifically, we randomly selected 100 samples from the unseen test set with a low b-value of 200 and a high b-value of 2000. These samples served as input for our proposed framework. For the t-SNE visualization, we extracted the feature maps of these selected cases from the bottleneck feature map of the baseline method.

# 3. Results

## 3.1 Patient Characteristics

This retrospective study contains 5,150 male cases with a median age of 65 years (IQR, 59-70 years), and the detailed patient and imaging characteristics are provided in Tab. 1.

Table 1: Patient and imaging characteristics

| Characteristic | Training (*n*=3458) | Test (*n*=1692) |
|---|---|---|
| Age (y)† | 65 (59-70) | 66 (60-70) |
| Manufacturer | | |
|   GE | 0 | 777 (43.8) |
|   Siemens | 3458 (100.0) | 915 (56.2) |
| Field strength (T) | | |
|   1.5 | 21 (0.6) | 24 (1.4) |
|   3 | 3437 (99.4) | 1668 (98.6) |
| Year of acquisition†† | 2014-2021 | 2014-2021 |
| T2w | | |
|   TR (*ms*)† | 5660 (4300-6500) | 4960 (4000-5660) |
|   TE (*ms*)† | 101 (101-104) | 104 (101-119) |
|   In-plane spacing (*mm*)† | 0.563 (0.469-0.625) | 0.391 (0.391-0.5) |
|   Slice thickness (*mm*)† | 3.6 (3.6-4) | 3.6 (3-4) |
| DWI | | |
|   TR (*ms*)† | 4000 (3600-5100) | 4025 (3003-4821) |
|   TE (*ms*)† | 73 (63-84) | 63 (59-63) |
|   In-plane spacing (*mm*)† | 1.625 (0.877-2) | 1.563 (0.938-2) |
|   Slice thickness (*mm*)† | 3.6 (3.6-4) | 3.5 (3-4) |
| PI-RADS category | | |
|   1, 2 | 1389 (40.1) | 804 (47.5) |
|   3 | 470 (13.6) | 176 (10.4) |
|   4 | 894 (25.9) | 419 (24.8) |
|   5 | 705 (20.4) | 293 (17.3) |

*Note.* — Unless otherwise noted, data are numbers of patients, with percentages in parentheses. T2w = T2-weighted imaging, DWI = diffusion-weighted imaging, TR = repetition time, TE = echo time, PI-RADS = Prostate Imaging Reporting and Data System, PZ = peripheral zone, TZ = transition zone.
† Data are medians, with first quartile to third quartile ranges in parentheses.
†† Year of acquisition is based on data for which acquisition dates are available

Table 2: Case-level AUC scores on unseen test sets. The 95% CI values are shown in Fig. S.1.

| | b-values | | Case-level AUC (*PIRADS* ≥ 3 / *PIRADS* ≥ 4) | | |
|---|---|---|---|---|---|
| Group | Low | High | Baseline | Generic | Proposed |
| 1 | 0 | 800, 1000 | 0.84 / 0.85 | 0.80 / 0.81 | 0.87 / 0.88 |
| 2 | 0 | 1200, 1400, 1500 | 0.74 / 0.83 | 0.87 / 0.88 | 0.86 / 0.92 |
| 3 | 50, 100 | 600, 700, 750 | 0.71 / 0.74 | 0.66 / 0.68 | 0.66 / 0.67 |
| 4 | 50, 100 | 800, 900, 1000 | 0.83 / 0.84 | 0.78 / 0.80 | 0.83 / 0.83 |
| 5 | 50, 100 | 1200, 1400, 1500, 1600 | 0.68 / 0.72 | 0.69 / 0.72 | 0.73 / 0.76 |
| 6 | 50, 100 | 2000 | 0.71 / 0.80 | 0.77 / 0.78 | 1.00 / 1.00 |
| 7 | 150, 200 | 800, 1000 | 0.90 / 0.95 | 0.80 / 0.85 | 0.95 / 0.97 |
| 8 | 150, 200 | 1500 | 0.88 / 0.92 | 0.83 / 0.86 | 0.96 / 1.00 |
| 9 | 150, 200 | 2000 | 0.49 / 0.50 | 0.61 / 0.65 | 0.76 / 0.77 |
| | Total | | 0.73 / 0.77 | 0.73 / 0.76 | 0.79 / 0.80 |

*Note.*— Baseline: a supervised learning pre-trained detection model (24); Generic model: retrained the baseline method using dataset with all possible b-value combinations. AUC = area under the receiver operating characteristic curve, PI-RADS = Prostate Imaging Reporting and Data System

## 3.2 Case-level Performance

Tab. 2 displays the case-level performance. For both PI-RADS≥3 and PI-RADS≥4 labels, the overall performance of the proposed model was significantly higher (p<.001) compared with other models. The baseline method was effective when test data b-values closely matched the training set, primarily in the reference domain (group 4), but struggled with OOD samples, as seen in group 9. Even with diverse b-values added to the training data, the generic model failed to differentiate across domains, resulting in unstable performance, especially in the commonly used reference domain. Our UDA method improved the results of the baseline method in all groups when b-values deviated from the standard and maintained consistent performance for the reference domain. The detailed results of the statistical analysis are shown in Fig. S.1. In addition, the overall AUC performance in different zones can be viewed in Tab. S.1.

Table 3: FROC results of baseline and proposed methods.

| Group | Methods (Baseline / Proposed method) | | | |
| --- | --- | --- | --- | --- |
| | TPR@FPp=0.75 ↑ | TPR@FPp=1 ↑ | FPp@TPR=0.65 ↓ | FPp@TPR=0.70 ↓ |
| 1 | 0.77 / 0.78 | 0.80 / 0.83 | 0.50 / 0.29 | 0.61 / 0.32 |
| 2 | 0.61 / 0.73 | 0.73 / 0.78 | 0.83 / 0.55 | 0.84 / 0.59 |
| 3 | 0.41 / 0.45 | 0.47 / 0.52 | 2.32 / 2.48 | 3.03 / 6.09 |
| 4 | 0.70 / 0.71 | 0.73 / 0.75 | 0.67 / 0.64 | 0.77 / 0.67 |
| 5 | 0.47 / 0.54 | 0.50 / 0.59 | 2.21 / 1.35 | 9.89 / 1.68 |
| 6 | 0.67 / 0.88 | 0.78 / 1.00 | 0.67 / 0.13 | 1.68 / 0.83 |
| 7 | 0.96 / 0.83 | 0.96 / 0.90 | 0.35 / 0.09 | 0.41 / 0.11 |
| 8 | 0.71 / 0.99 | 0.83 / 1.00 | 0.33 / 0.10 | 7.67 / 0.18 |
| 9 | 0.20 / 0.57 | 0.23 / 0.60 | 5.19 / 1.49 | 5.19 / 2.15 |
| Total | 0.45 / 0.64 | 0.53 / 0.68 | 1.64 / 0.78 | 2.31 / 1.11 |

*Note.* — ↑ and ↓ represent "higher is better" and "lower is better", respectively. TPR and FPp denote true positive rate and false positive per patient, respectively. In details, the true positive rate (TPR=0.75 and 1) and false positives per patient (FPp=0.65 and 0.70) are used for the evaluation. FROC = free-response receiver operating characteristic

### 3.3 Lesion-level Performance

FROC values representing the FP to TP ratio are shown in Tab. 3. The proposed method outperformed the baseline in most domains except for the FP rate in group 3 and the TP rate in group 7. Fig. 4 shows four typical domains with the highest testing samples. For standard b-values, the baseline method had higher AUC scores; but, the proposed method demonstrates higher performance of FROC, particularly when b-values diverged from the standard.

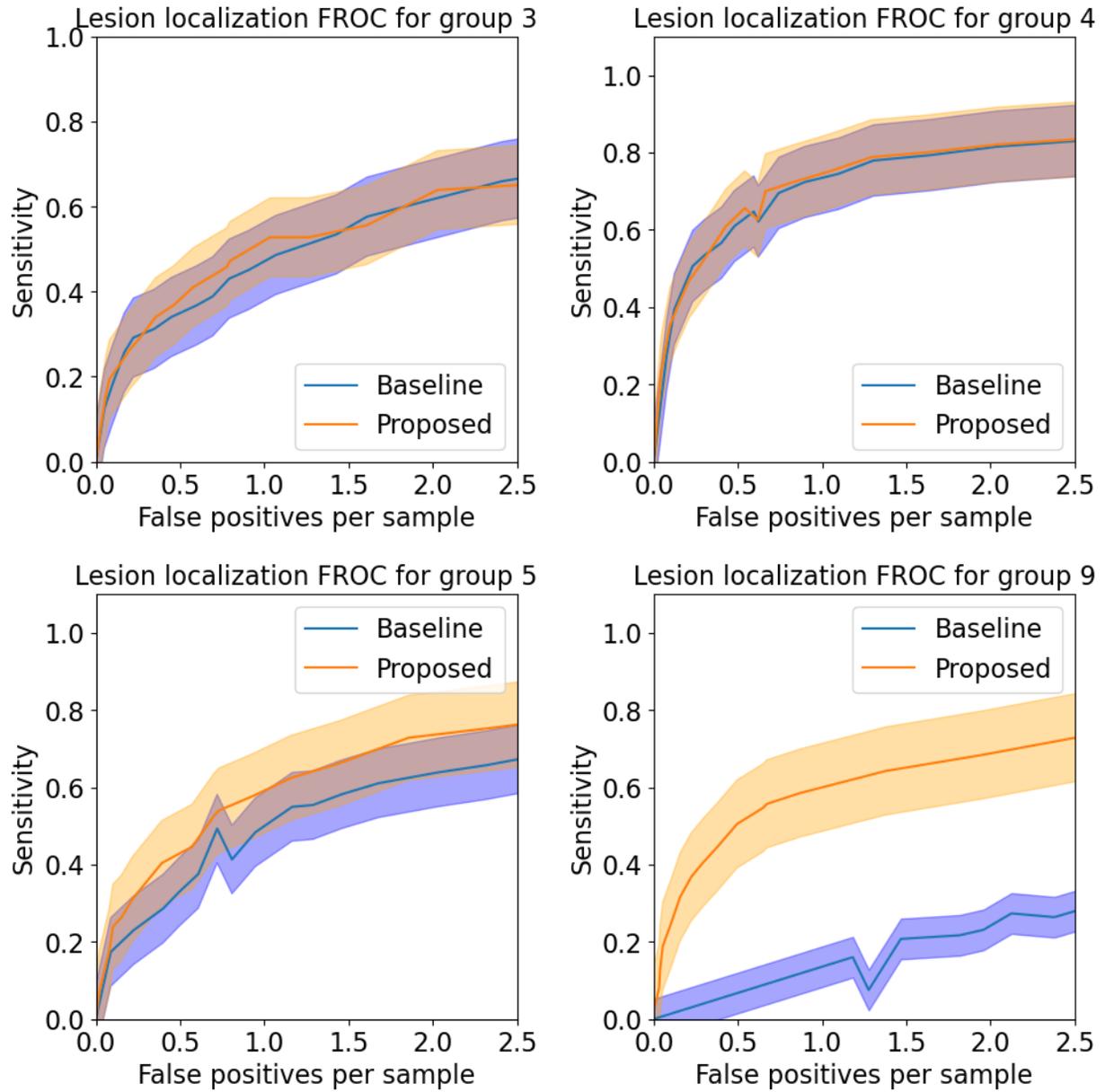

*Figure 4: FROC curves of comparative methods. Blue and orange represent the baseline and proposed methods, respectively. The 95% confidence intervals for each are shaded. The selected groups are common actual b-value imaging settings and fall into three categories: near, within, and distant from the standard b-value range. FORC=free-response receiver operating characteristic.*

## 3.4 Quality of Generated Image

The proposed framework was assessed in terms of the image quality of generated DWI B-2000 images through comparison with paired DWI B-2000 images in the reference domain. The PSNR, MSE, and SSIM results are provided in Tab. 4. In the analysis of both b-value pairs, the generated images had higher PSNR, lower MSE and higher SSIM. The results indicate that the DWI B-2000 images generated by our method are more similar to the reference domain images than the original target domain images.

Fig. 5 displays original bpMRI images, generated DWI images and detection heatmaps of 4 example cases (2 positive and 2 negative). In each group, one case is from the reference domain and one case is from the target domain with, b-values far from the PI-RADS guideline recommendation. For cases in the reference domain, no obvious changes were observed in the generated images, and the predictions were similar by using original or generated images. For cases not in the reference domain, a significant improvement in image quality was observed for the generated images. The predictions were also more accurate when compared with the reference standard annotations.

Table 3: Image similarity comparison of original and computed DWI b-2000 images, presented as PSNR / MSE ($\times 10^{-3}$) / SSIM.

| b-values | | DWI | |
|---|---|---|---|
| Low | High | Original | Generated |
| 150 | 1500 | 27.65 / 1.85 / 0.849 | 32.89 / 0.74 / 0.889 |
| 200 | 2000 | 20.21 / 10.1 / 0.625 | 31.21 / 1.11 / 0.840 |

*Note.* — The results are calculated by comparing original and generated images with their corresponding images in the reference domain. DWI = diffusion-weighted imaging, MSE = mean square error, PSNR = peak signal-to-noise ratio, SSIM = structural similarity index measure

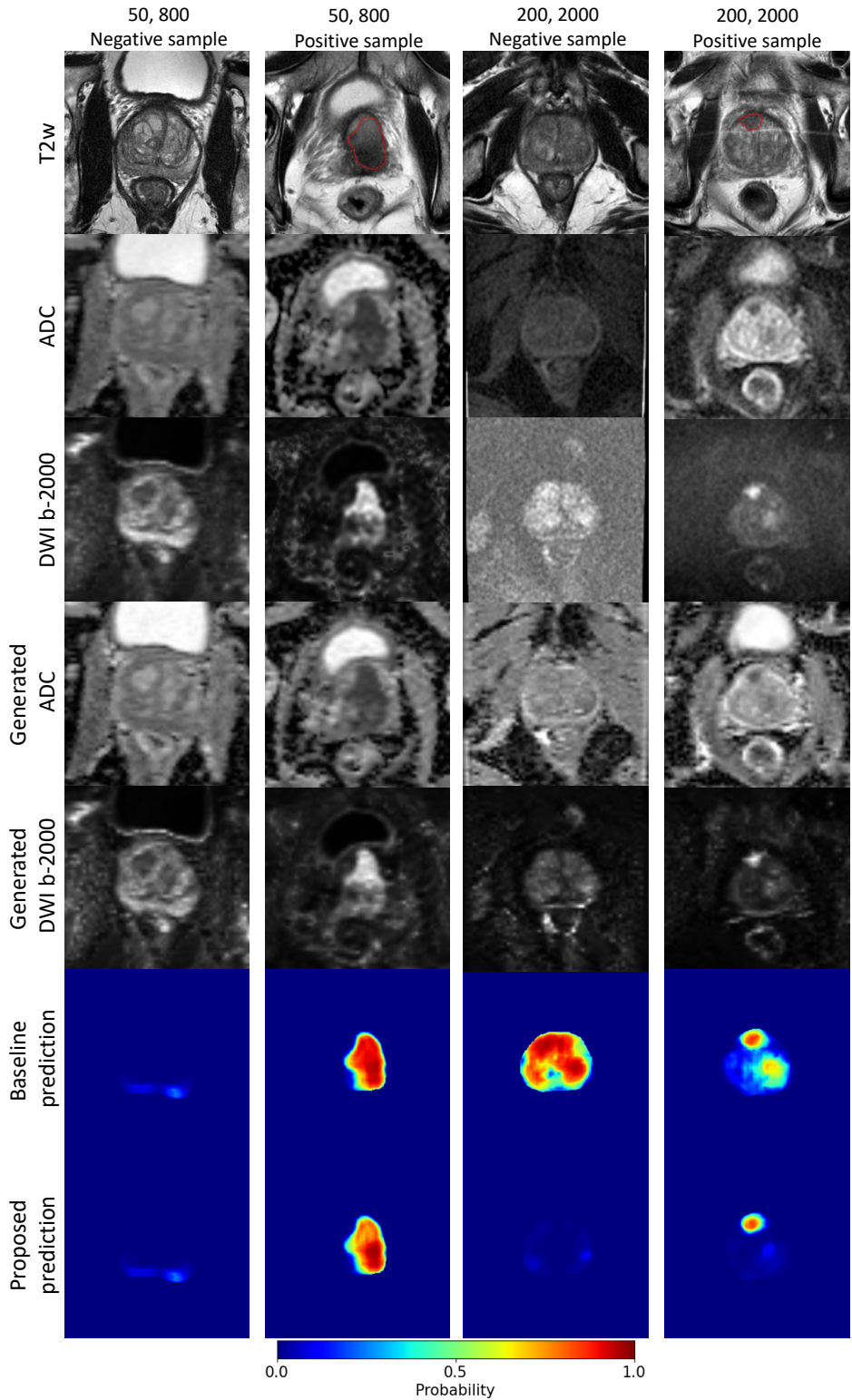

*Figure 5: Qualitative results of axial plane from four example samples are presented based on their reference standard labels and b-values, which are indicated at the top of each column and represented as "low b-value, high b-value". The type of image (T2w and DWI) for each sample is labeled on the left side of the figure. Red contours outline the reference standard lesion annotations. All images in the same row are displayed using the same window center and width.*

## 3.5 t-SNE Visualization

To better visualize the relationship between reference domain images, original target domain images and generated images, we applied the t-SNE algorithm to reduce the bottleneck layer features of the PCa detection network to a 2-dimensional representation. For neat visualization, 100 cases whose original DWI acquisitions had low b-value=200 and high b-value=2000 were selected for this analysis. The original and generated DWI images were used as input of the detection network. Another 100 cases from the PROSTATEx Challenge dataset (32) were selected as the reference domain (i.e. low b-value=50 and high b-value=800) cases for comparison. The 2-dimensional plot of the bottleneck layer features are displayed in Fig. 6. The generated DWI images using the proposed method formed a tighter cluster compared with the original target domain image. Moreover, the generated images aligned more closely with the reference domain data, which indicates a higher similarity in the latent space of the detection network.

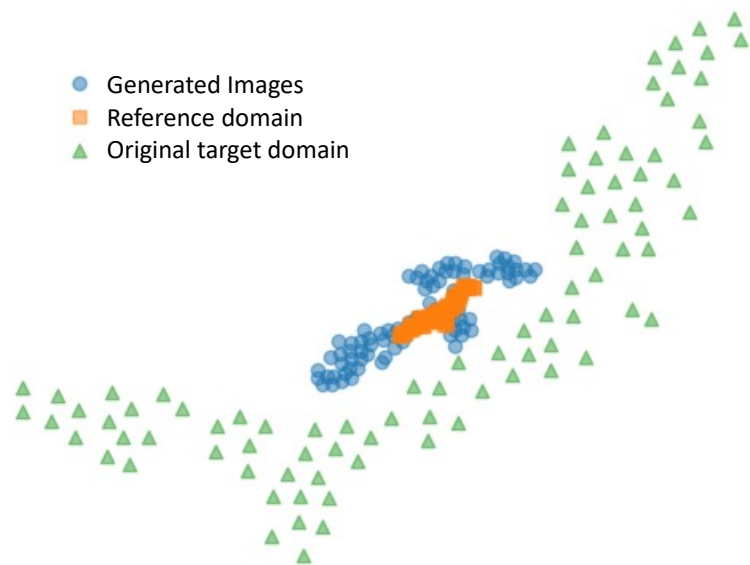

*Figure 6: t-SNE visualization based on the feature map of the pre-trained detection network at bottleneck layer. Orange squares represent inputs from reference domain images, green triangles denote inputs using original target domain images, and blue dots correspond to inputs of images generated by the proposed method. t-SNE=t-distributed Stochastic Neighbor Embedding.*

## 4. Discussion

We proposed a novel UDA method with a unified model to solve practical common issues, specifically domain shift and label availability, for PCa lesion detection. Only a unified model is used in our framework for multi-domain mapping instead of multiple networks being trained as in typical UDA methods. To achieve better performance of a unified model in multi-domain scenarios, we proposed and employed a dynamic filter to leverage domain information. When benchmarked against other methods using a large-scale, multi-site dataset comprising 5,150 cases (14,191 samples), our approach consistently demonstrated an enhanced capability to perform more accurate PCa detection.

To demonstrate the feasibility for practical use, this study was conducted on a large-scale dataset with different imaging protocols, where the heterogeneous domain shifts are present and pose a challenge to achieve a consistent performance. The proposed method leverages information from the entire dataset, notably unlabeled data, to reduce the annotation effort, which is usually a burden for large datasets. Importantly, the proposed method can seamlessly be integrated into any pre-trained PCa detection framework and can be used as an image adapter at the upstream level to reduce discrepancies between domains. The method overall improved the generalizability of downstream PCa detection models. Importantly, there is no need to retrain or modify the network for new target data, making the method suitable for a variety of medical image applications.

To our knowledge, no prior study has explored the domain shift in PCa detection using ADC and DWI high b-value images, especially given the various domains in our study. We validated the common practical solutions as part of our experimental contributions. Although such methods are not the optimal solutions, important findings emerged: (1) using the original test image is preferable if its b-values closely align with training set, and (2) retraining a generic model could produce unpredictable results due to its broad adaptability and limitations in specific learning. These findings offer valuable insights for future studies, especially those targeting clinical applications.

Several existing studies have tried to address these prevalent practical challenges by producing more consistent DWI images. For instance, in (9), the authors suggested recalculating ADC maps and high b-value images at a fixed 2000 $sec/mm^2$ rather than using the originally acquired images. However, this method cannot avoid the diffusion kurtosis effect if the acquired DWI uses a high b-value over 1500 $sec/mm^2$. The proposed method with an image-to-image technique effectively translates OOD target samples into the style of reference domain. Resulting generated ADC and DWI b-2000 images are similar to the real reference domain image both at image-level and latent-level. The most pronounced improvements occur when the high b-value deviates farther from the standard range. Additionally, when high b-values are within the reference domain, our method boosts performance for low b-values that are OOD samples. Comprehensive detection results indicate that high b-values influence domain discrepancy more than low b-values.

In this study, we introduce the dynamic filter, which can be treated as a domain indicator and plugged it into any generator to leverage meta-information. The proposed dynamic filter generates conditional parameters according to the corresponding meta-information to differentiate domains. This is unlike encoding meta-information as one-hot vectors, which often requires a codebook to rigidly encode the corresponding relationship. Such an encoding approach might not be ideal for large-scale studies involving multiple domains. Additionally, the codebook would need adjustments whenever a new combination of b-values emerges. In contrast, we designed an effective yet straightforward strategy that directly uses b-values as input. This not only retains the original meta-information but also simplifies the process, making it adaptable to arbitrary b-value combinations.

Our study had some limitations. Our approach is limited by utilizing only b-values from meta-information to provide domain information. This constraint may contribute to suboptimal performance, especially when high b-values closely resemble those of the reference domain. It is crucial to acknowledge this limitation while also recognizing the potential expansion of our proposed domain adaptation method to include T2w images in future iterations. Subsequent efforts will involve incorporating additional meta-information, such as field strength, sequence selection, and the number of averages, to more comprehensively quantify domain shifts resulting from diverse acquisition protocols. Another potential contributing factor to the occasional unsatisfactory performance of our proposed method is the restricted number of training samples available for specific b-value settings, as evident in the comparatively lower performance of group 3, comprising only 53 training samples. Future endeavors will concentrate on manifold learning within the latent space of meta-information, offering a continuous representation of b-value variations and enhancing generalizability to unseen b-value settings. Moreover, as our proposed UDA framework demonstrates its feasibility, subsequent work will involve developing additional image synthesis models based on our current approach. A comparative analysis of widely used synthesis models will be a crucial step forward. Additionally, recent developments in network architectures like nnU-Net and its variants have proven effective for PCa detection tasks (33, 34). The detection network could be redesigned using similar architectures to better adapt to synthesized images, potentially enhancing detection accuracy. As an early-stage study, we observed a significant improvement in overall performance with the proposed method, particularly when the high b-value deviates farther from the standard range. These improvements indicate that the proposed method may provide higher quality images and more accurate detection results, facilitating the interpretation time for radiologists, particularly those who are less experienced, and potentially increasing inter-reader agreement. However, the performance is constrained by the imbalanced cases of b-value distribution, not only between training and test sets for certain groups but also within different groups in the training set. In practice, the test sample from such groups may directly feed into the detection network. Future work will aim to improve the accuracy across diverse data distribution using cost-sensitive learning.

In conclusion, our unified model-based UDA method for multi-site PCa detection showed marked improvement compared with the baseline model on a large dataset, especially outside the reference domain. To the best of our knowledge, this is the first large-scale study exploring the impact of b-value properties on ADC and DWI b-2000 images with the aim of improving detection outcomes for a multi-domain scenario. It also paves the way for future research, potentially inspiring further studies in this field.

**Disclaimer**. The concepts and information presented in this paper are based on research results that are not commercially available. Future commercial availability cannot be guaranteed.

# Supplementary Materials

## S.1 Details of Annotation Process

The annotation process for the voxel-level prostate lesion segmentation and Prostate Imaging Reporting and Data System (PI-RADS) score was done as follows. Initially, we collated all clinical radiology reports, each accompanied by the respective PIRADS score for individual lesions, along with their corresponding lesion annotations. The raw clinical annotations exhibited variations, comprising landmarks indicating the lesion, bounding boxes around the lesion, or single contours on at least one slice to guide annotators regarding the location of the lesion of interest. In the subsequent step, annotators were tasked with delineating a complete 3D mask of the lesion based on the original annotations. All annotators underwent training provided by radiologists with Doctor of Medicine degrees and residency in Radiology. Supervised by radiologists, annotators received guidance throughout the process, addressing any uncertainties that arose. In the third step, 3D annotations from annotators underwent review and necessary corrections by radiologists within the annotation team. Radiologists possessed the authority to overrule annotators' annotations when deemed necessary. In the final step, all annotations and corresponding clinical reports underwent meticulous review by an expert radiologist with five years of experience in radiology, specializing in prostate MRI examinations. As part of the annotation, we relied on the clinical assignment of PI-RADS as determined by the original clinical site. The primary objective of the annotation process was to harmonize and extend lesion contouring across slices, providing a 3D mask for each identified lesion. The expert radiologist made minimal changes to the original PI-RADS assignment (<1%) during case reviews to ensure consistency in applying the same PI-RADS criterion across all datasets. All annotations were performed using an internally customized tool where anonymized MRI DICOM series were loaded. It's important to note that only lesions with PI-RADS scores of 3 and above were annotated in this study.

**S.2 Image Preprocessing**

Initially, the original T2-weighted (T2w) and diffusion-weighted imaging (DWI) series were extracted from the raw DICOM files. We employed a voxel-wise logarithmic extrapolation of the fitted signal decay curves to compute the apparent diffusion coefficient (ADC) map and new DWI volumes at b-values of 0 and 2000 $sec/mm^2$ which are denoted as DWI b-0 and b-2000 images. The same process was repeated for each pair of low and high b-values if the case has more than two b-values. Next, whole prostate gland segmentation was performed on T2w volumes using the method presented in (35). Subsequently, all DWI volumes were aligned to the T2w volumes using rigid registration. This alignment was meticulously verified through visual inspection. Additionally, volumes were center cropped and resampled to an image dimension of $240 \times 240 \times 30$ and a voxel spacing of $0.5 \times 0.5 \times 3\ mm^3$. Finally, we normalized all volumes to facilitate the training process. For T2w volumes, we linearly normalized to the range [0, 1] based on the 0.05 and 99.95 percentiles of their intensities. Given that ADC volumes represent quantitative parametric maps, they were normalized using a constant factor of 3000. For computed DWI b-2000 volumes, we first normalized them by a factor obtained as median intensity within the prostate gland region of the corresponding DWI b-0 volumes, and then normalized by a constant value to linearly map the approximate range to [0, 1].

## S.3 Training Details of Proposed Framework

| Algorithm 1: Training process of DWI generator → $G(\theta)$ |
|---|
| **Input**: DWI b-2000 → $x$ and its meta-information → $m = (low-b, high-b)$ |
| 1  Load pre-trained detection network → $D$ and its training set → $D_{train}$; |
| 2  **while** $epoch \leq 100$ **do** |
| 3  $\quad$ **if** $x \in D_{train}$ **then** |
| 4  $\quad\quad$ $L_{syn} = L_{det} + L_{CUT}$; |
| 5  $\quad\quad$ **if** $50 \leq low-b \leq 100$ and $800 \leq high-b \leq 1000$ **then** |
| 6  $\quad\quad\quad$ $L_{syn} = L_{det} + L_{CUT} + L_{consistency}$; |
| 7  $\quad\quad$ **end** |
| 8  $\quad$ **else** |
| 9  $\quad\quad$ $L_{syn} = L_{CUT}$; |
| 10 $\quad$ **end** |
| 11 $\quad$ Update $\theta$; |
| 12 **end** |

*Note.* — DWI = diffusion-weighted imaging

## S.4 Statistical Analysis of Case-level AUC

The comparison of case-level area under the receiver operating characteristic curve (AUC) between baseline and proposed methods are displayed in Fig. S.1.

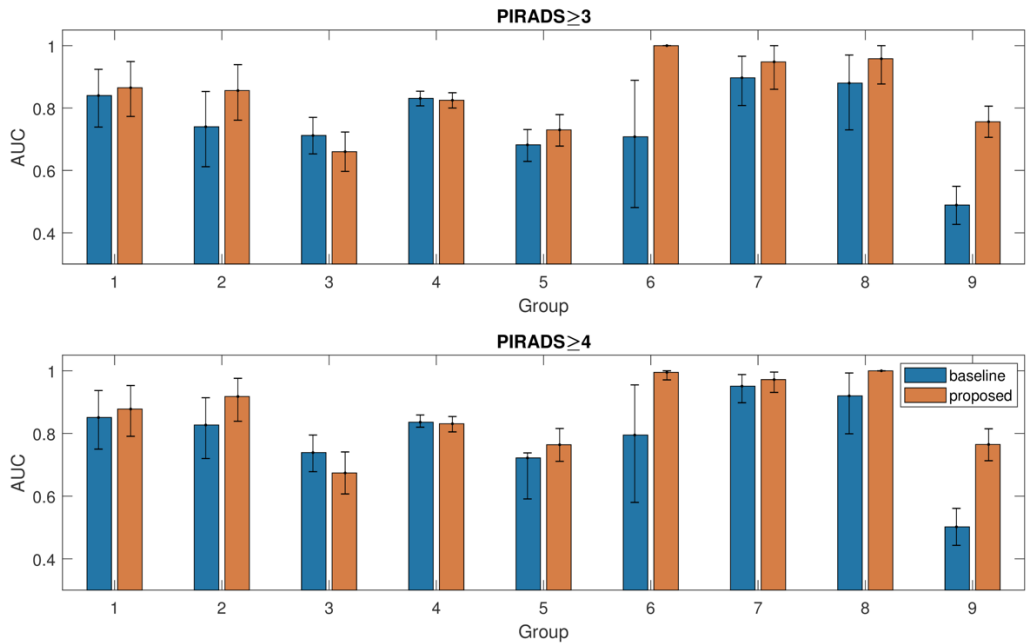

*Figure S.1: Comparison of case-level area under the receiver operating characteristic curve (AUC) between baseline and proposed methods. Error bars indicate the 95% confidence interval (95% CI) using a bootstrapping approach. The AUC was calculated based on bootstrap resampling of test samples for each group and repeated for 2000 times. The 2.5th and 97.5th percentile of the bootstrapped AUC distribution was used for an estimation of the 95% CI.*

## S.5 Zone Analysis of Case-level AUC

Table S.1: The overall AUC performances (PI-RADS≥3) across various zones are presented as AUC [lower 95% CI, higher 95% CI]. Negative sample number is 1212.

|  | Methods | | |
|---|---|---|---|
| Zone | Baseline | Generic | Proposed method |
| PZ ($n = 555$) | 0.69 [0.67, 0.72] | 0.67 [0.64, 0.70] | 0.73 [0.71, 0.76] |
| TZ ($n = 349$) | 0.79 [0.77, 0.82] | 0.80 [0.77, 0.82] | 0.82 [0.79, 0.85] |
| Both ($n = 277$) | 0.76 [0.73, 0.79] | 0.73 [0.70, 0.77] | 0.80 [0.77, 0.83] |

*Note.* — n denotes sample number, PI-RADS= Prostate Imaging Reporting and Data System, AUC=area under the receiver operating characteristic curve, CI=confidence interval.

We present the overall AUC performance across all b-value groups for cases with lesions in different zones. In this analysis, cases were categorized into a specific zone if 80% of their lesions were within that area; otherwise, they were labeled as "both". Given the absence of prostate gland annotations, we generated zone masks using the gland segmentation model outlined in Section S.1. The observed improvement is statistically significant (p<0.05) for all zone categories when com-paring our proposed method with other approaches. Notably, cases with lesions in the peripheral zone (PZ or "both") exhibited a higher margin of performance improvement compared to cases with transitional zone lesions only. This finding is in accordance with the PI-RADS guideline, high-lighting the influential role of DWI images in determining the PI-RADS assessment for lesions in the peripheral zone.

## S.6 Ablation Study

The compared methods are: (1) CUT (Contrastive learning for Unpaired image-to-image Translation (27)): our solution (see Fig. 1(c)) to domain shift that utilizes CUT to achieve unpaired image-to-image (I2I) translation; (2) CUTe: Form CUT into an end-to-end workflow; (3) CUTd: CUT with dynamic filter; (4) CUTd+e: Form CUTd into an end-to-end workflow; and (5) the proposed method: Additional consistency loss added on the CUTd+e approach. The details of the compared methods can be viewed in Fig. S.2. For a fair comparison, the same network architecture is used for the detection model and generator, except for the additional controller for the methods with dynamic filters. All unsupervised domain adaptation (UDA) results are produced from the baseline method (24).

Tab. S.2 details the ablation study on the DWI generator, showing that UDA with image-to-image translation improves PCa detection for some groups. Performance in CUT and CUTe was unstable, particularly in group 4. By considering domain information with CUTd, improvements were observed in groups 1, 2, 6, 7, 8, and 9, highlighting the efficacy of the dynamic filter in varied domains. Building on CUTd, an end-to-end framework (CUTd+e) was tested. However, the image-to-image mapping of the generator might be influenced by the pre-trained detection model from various domains, potentially causing misalignment of generated images and leading to reduced accuracy. To address this, we introduced a consistency loss in our proposed method to retain details of input samples with b-values from the standard PI-RADS range to counter biases from the pre-trained supervised learning (SL) model. The qualitative results can be viewed in Fig. S.3.

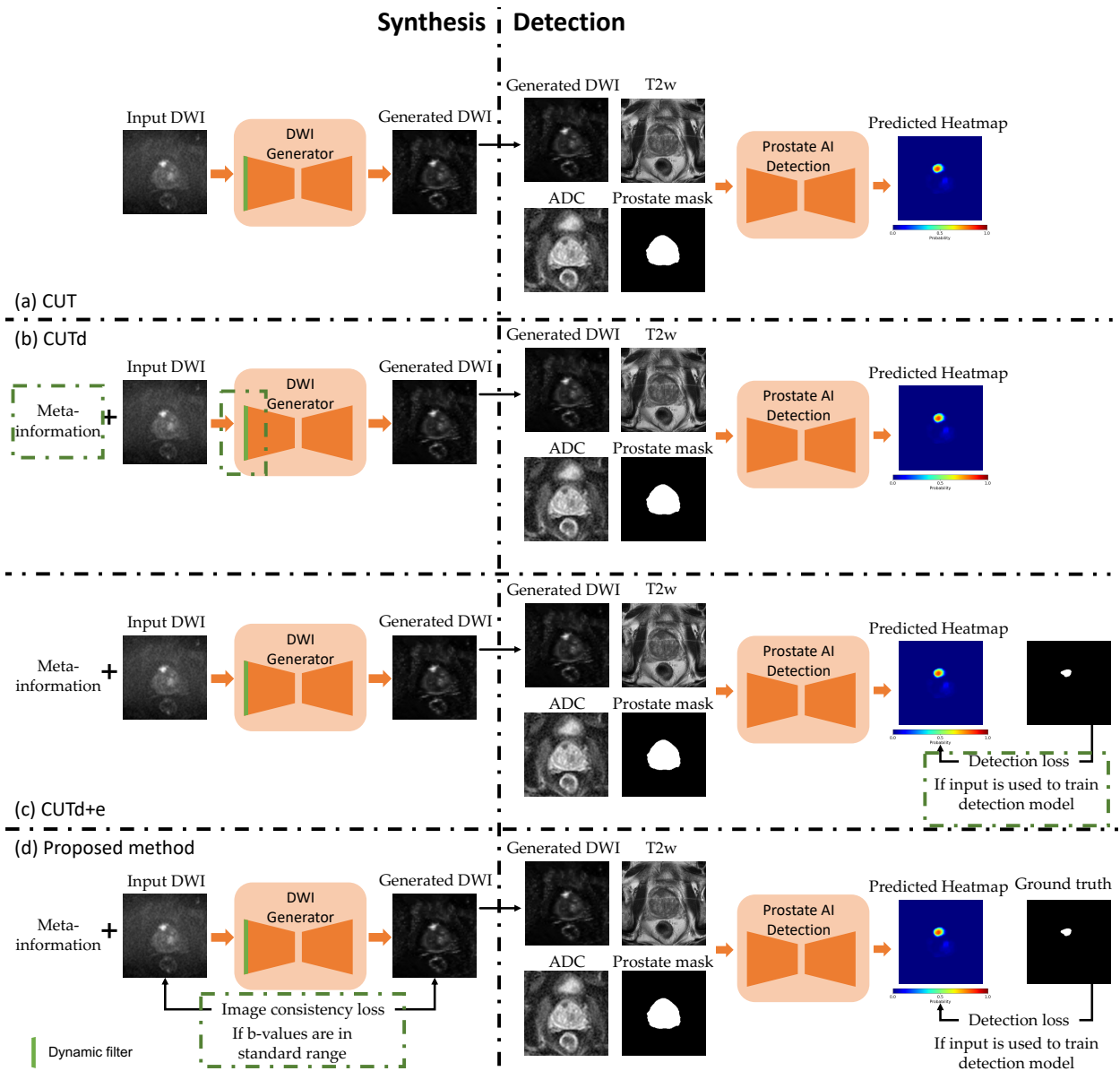

*Figure S.2: Illustrations of diffusion-weighted imaging (DWI) generators among the compared methods are as follows: (a) CUT, (b) CUTd, (c) CUTd+e, and (d) proposed methods. The green boxes highlight the improvements made from the previous step. CUTe is excluded for brevity, and its improvements can be observed between (b) and (c).*

Table S.2: Quantitative AUC score of ablation studies on DWI b-2000 images only.

| | Ablation study of DWI generator (*PIRADS* ≥ 3 / *PIRADS* ≥ 4) | | | | | |
|---|---|---|---|---|---|---|
| Group | Baseline | CUT | CUTe | CUTd | CUTd+e | Proposed |
| 1 | 0.84 / 0.85 | 0.80 / 0.83 | 0.76 / 0.78 | 0.83 / 0.85 | 0.82 / 0.83 | 0.85 / 0.87 |
| 2 | 0.74 / 0.83 | 0.78 / 0.78 | 0.77 / 0.87 | 0.74 / 0.83 | 0.78 / 0.88 | 0.83 / 0.92 |
| 3 | 0.66 / 0.74 | 0.67 / 0.70 | 0.64 / 0.68 | 0.59 / 0.59 | 0.69 / 0.72 | 0.66 / 0.67 |
| 4 | 0.83 / 0.84 | 0.81 / 0.81 | 0.80 / 0.80 | 0.79 / 0.78 | 0.81 / 0.82 | 0.83 / 0.83 |
| 5 | 0.68 / 0.72 | 0.69 / 0.73 | 0.67 / 0.71 | 0.64 / 0.65 | 0.68 / 0.73 | 0.71 / 0.75 |
| 6 | 0.71 / 0.80 | 0.82 / 0.92 | 0.88 / 0.94 | 0.83 / 0.84 | 0.90 / 0.98 | 0.85 / 0.80 |
| 7 | 0.90 / 0.95 | 0.90 / 1.00 | 0.89 / 0.86 | 0.92 / 0.90 | 0.89 / 0.93 | 0.92 / 0.95 |
| 8 | 0.88 / 0.92 | 0.83 / 0.91 | 0.90 / 0.92 | 0.94 / 0.94 | 0.83 / 0.92 | 0.91 / 0.96 |
| 9 | 0.49 / 0.50 | 0.67 / 0.70 | 0.63 / 0.67 | 0.68 / 0.70 | 0.67 / 0.71 | 0.70 / 0.73 |

*Note.* — "e" notes the end-to-end training with supervision by the pre-trained detection network. "d" represents the usage of the dynamic filter. AUC=area under the receiver operating characteristic curve, PI-RADS= Prostate Imaging Reporting and Data System

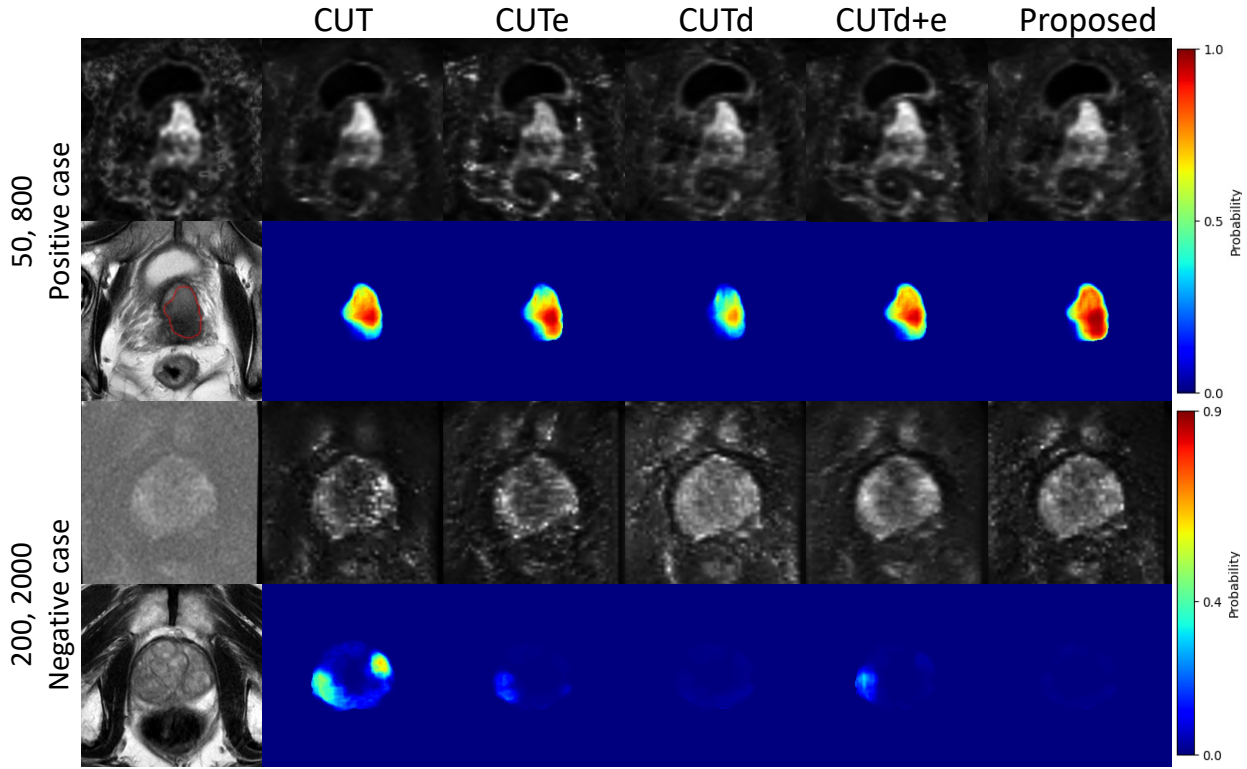

*Figure S.3: Qualitative results of various UDA methods from ablation studies on DWI B-2000 images. The figure displays B-2000 images alongside their corresponding prediction heatmaps.*